\documentclass[twocolumn,apl,epsf,superscriptaddress]{revtex4}
\usepackage{amsmath}
\usepackage{amsfonts}
\usepackage{epsfig}
\usepackage{epsf}
\usepackage{array}
\usepackage{color}
\usepackage{ulem}

\begin{document}

\title{Accuracy of the microcanonical Lanczos method to compute real-frequency
dynamical spectral functions of quantum models at finite temperatures}
\altaffiliation{
Copyright  notice: This  manuscript  has  been  authored  by  UT-Battelle, LLC under Contract No. DE-AC05-00OR22725 with the U.S.  Department  of  Energy.   
The  United  States  Government  retains  and  the  publisher,  by  accepting  the  article  for  publication, 
acknowledges  that  the  United  States  Government  retains  a  non-exclusive, paid-up, irrevocable, world-wide license to publish or reproduce the published form of this manuscript, 
or allow others to do so, for United States Government purposes.  
The Department of Energy will provide public access to these results of federally sponsored  research  in  accordance  with  the  DOE  Public  Access  Plan 
(http://energy.gov/downloads/doe-public-access-plan)}

\author{Satoshi Okamoto}
\altaffiliation{okapon@ornl.gov}
\affiliation{Materials Science and Technology Division, Oak Ridge National Laboratory, Oak Ridge, Tennessee 37831, USA}
\author{Gonzalo Alvarez}
\affiliation{Center for Nanophase Materials Sciences, Oak Ridge National Laboratory, Oak Ridge, Tennessee 37831, USA}
\affiliation{Computational Science and Engineering Division, Oak Ridge National Laboratory, Oak Ridge, Tennessee 37831, USA}
\author{Elbio Dagotto}
\affiliation{Materials Science and Technology Division, Oak Ridge National Laboratory, Oak Ridge, Tennessee 37831, USA}
\affiliation{Department of Physics and Astronomy, The University of Tennessee, Knoxville, Tennessee 37996, USA}
\author{Takami Tohyama}
\affiliation{Department of Applied Physics, Tokyo University of Science, Tokyo 125-8585, Japan}

\begin{abstract}
We examine the accuracy of the microcanonical Lanczos method (MCLM) developed by 
Long, {\it et al.} [Phys. Rev. B {\bf 68}, 235106 (2003)] 
to compute dynamical spectral functions of 
interacting quantum models at finite temperatures. 
The MCLM is based on the microcanonical ensemble, which becomes exact in the thermodynamic limit.  
To apply the microcanonical ensemble at a fixed temperature, one has to find energy eigenstates with the energy eigenvalue  
corresponding to the internal energy in the canonical ensemble. 
Here, we propose to use thermal pure quantum state methods by 
Sugiura and Shimizu [Phys. Rev. Lett. {\bf 111}, 010401 (2013)] 
to obtain the internal energy. 
After obtaining the energy eigenstates using the Lanczos diagonalization method, 
dynamical quantities are computed via a continued fraction expansion, 
a standard procedure for Lanczos-based numerical methods. 
Using one-dimensional antiferromagnetic Heisenberg chains with $S=1/2$, 
we demonstrate that the proposed procedure is reasonably accurate even for relatively small systems. 
\end{abstract}


\maketitle

\date{\today }


\section{Introduction}

Strongly correlated electron systems have been one of the main areas of research in condensed matter physics 
because of a variety of emergent phenomena arising from many-body effects \cite{Anderson1972,Imada1998}. 
The many-body correlations in the strong coupling limit are notoriously difficult to handle because 
the single-particle approximation is not applicable. 
Thus, a significant amount of effort has been devoted to developing theoretical or numerical techniques to 
solve interacting Hamiltonians unperturbatively. 

The Lanczos algorithm \cite{Lanczos1950} is an exact method to study many-body effects numerically, 
by {\it partially} and iteratively diagonalizing the Hamiltonian matrix to obtain one or more eigenstates. 
This method has been used for the study of high-$T_c$ cuprate ground states and excited states alike \cite{Dagotto1994}. 
The Lanczos method was readily extended to deal with finite  temperature, called the finite temperature Lanczos method (FTLM) 
\cite{Jaklic1994}. 
Additional improvements were later proposed for the FTLM \cite{Aichhorn2003,Munehisa2014,Munehisa2015}. 

A fundamental problem of Lanczos-based procedures is that the dimension of the many-body Hamiltonian matrices grows exponentially 
with the system size. 
The density-matrix renormalization-group (DMRG) method avoids this exponential growth by truncating the less important states in a systematic way 
\cite{White1992,White1993}. 
Similar to the Lanczos-based methods, DMRG has been extended to deal with finite temperature \cite{Feiguin2005,Sota2008,Tiegel2014}. 

Most of the numerical procedures addressing finite temperature are based on the canonical ensemble. 
Recently, an alternative approach was proposed based on the microcanonical ensemble, called the microcanonical Lanczos method (MCLM) \cite{Long2003,Zotos2006,Prelovsek2013}. 
The MCLM is believed to become accurate in large systems close to the thermodynamic limit 
because the microcanonical ensemble and the canonical ensemble become equivalent in this limit  \cite{Landau}. 
Moreover, unlike FTLM, it is not necessary to average over many energy eigenstates with the appropriate Boltzmann weight.  
Thus, the computational cost of the MCLM and the zero-temperature Lanczos method is of the same order. 

Similar to but distinct from the exact diagonalization methods, 
other approaches have been proposed, called thermal pure quantum state (TPQ) methods \cite{Sugiura2012,Sugiura2013}. 
We write ``TPQ methods'' because there are microcanonical and canonical versions. 
The TPQ methods do not require diagonalization of the Hamiltonian matrices. 
Instead, TPQs are constructed iteratively starting from a random initial vector. 
It has been shown that the TPQ methods are 
able to handle larger systems than exact diagonalization methods and 
are able to compute the static quantities of many-body Hamiltonians accurately \cite{Sugiura2012,Sugiura2013}. 
However, it is not known if the TPQ methods could be extended to compute the dynamical properties or real-frequency properties of 
many-body Hamiltonians,  
while the real-time dynamics of many-body models at finite temperatures have been investigated using similar techniques \cite{Jin2010,Steinigeweg2014,Jin2015}. 

This paper examines the applicability of the MCLM method developed in Ref.~\cite{Long2003} 
to small size systems. 
This method needs the internal energy of the system of interest at a given temperature.
For this purpose, we use the TPQ method developed in Ref.~\cite{Sugiura2013}. 
Because TPQ needs no diagonalization, and because the number of diagonalization processes that {\it are} needed for the rest is very small,
the computational cost of the whole process is similar to that of the zero-temperature Lanczos calculations. 
This makes the current procedure very attractive, but the system size is still limited by the exponential growth of the basis set or Hilbert space. 
Thus, in order to apply the new method to larger systems, one may have to combine the current procedure with, for example, DMRG, 
which can truncate the less important states systematically. 

The rest of the paper is organized as follows:
In Sec.~\ref{sec:MM}, we describe the MCLM and introduce a model system. 
In Sec.~\ref{sec:results}, we present the numerical results. 
In Sec.~\ref{sec:summary}, we summarize our results and present discussions on issues and possibilities of TPQ 
to compute dynamical quantities from many-body Hamiltonians. 
We also briefly mention an idea to combine the MCLM and DMRG. 

\section{Methodology and model}
\label{sec:MM}

To begin with, we briefly describe the MCLM proposed in Ref.~\cite{Long2003}. 
This method is based on the physical principle that the microcanonical ensemble and the canonical ensemble give the same results in the thermodynamic limit. 
In this limit, the energy distribution function $\rho(\varepsilon)$ in the microcanonical ensemble is sharply peaked at $\varepsilon=\lambda$, with 
$\lambda$ corresponding to the average internal energy $E=\langle \hat H \rangle$ in the canonical ensemble at a given $T$. 
Here, $\hat H$ is an interacting Hamiltonian describing the system of interest. 

Suppose one could find eigenstates $| \psi_\lambda \rangle$ of an interacting Hamiltonian $\hat H$, 
i.e., $\hat H| \psi_\lambda \rangle = \lambda | \psi_\lambda \rangle $. 
Then, the expectation value of an operator, say $\hat O$, is given by $\langle \hat O \rangle = \langle \psi_\lambda| \hat O | \psi_\lambda \rangle$. 
If there is more than one eigenstate with the same energy $\lambda$, 
one has to average over these eigenstates. 
Dynamical quantities can be also computed in the same manner. 
Here, we consider a dynamical function defined by 
\begin{equation}
S(\vec q, \omega) = \frac{1}{2 \pi} \int_{-\infty}^\infty dt e^{i \omega t} \langle \hat X_{\vec q} (t) \hat X_{-\vec q} (0) \rangle ,
\end{equation}
where $\hat X_{\vec q} (t)$ is $\hat X_{\vec q} (t) = e^{i \hat H t} X_{\vec q} \, e^{-i \hat H t}$ in the Heisenberg representation. 
Noticing that $| \psi_\lambda \rangle$ is an energy eigenstate, one obtains 
\begin{eqnarray}
\hspace{-1em}
S(\vec q, \omega)
\!\!\! &=&\!\!\! \frac{1}{2 \pi} \int_{-\infty}^\infty dt e^{i (\omega + \lambda) t} 
\langle \psi_\lambda | \hat X_{\vec q} \, e^{-i \hat H t} \hat X_{-\vec q} | \psi_\lambda \rangle \nonumber \\
&=& \!\!\! - 
\frac{1}{\pi} \, {\rm Im}
\langle \psi_\lambda | \hat X_{\vec q} \Bigl[\omega+i \eta + \lambda - \hat H \Bigr]^{-1} \! \hat X_{-\vec q} | \psi_\lambda \rangle ,
\label{eq:sqw} 
\end{eqnarray}
where $i \eta$ is a small imaginary number introduced to approximate a $\delta$ function by a Lorentzian. 
As in a standard Lanczos procedure, one can compute this quantity using a continued fraction expansion (CFE) \cite{Dagotto1994}
starting from a new vector $\hat X_{-\vec q} | \psi_\lambda \rangle$.  
The above procedures are exactly the same as those at zero temperature except that expectation values are not taken in the ground state 
$| \psi_0 \rangle$ but in excited states $| \psi_\lambda \rangle$. 

We now address a technical problem in obtaining eigenstates with an arbitrary $\lambda$. 
As described in Ref.~\cite{Long2003}, one can consider a new operator 
\begin{equation}
\hat K_\lambda = (\hat H -\lambda)^2 
\end{equation}
and perform a Lanczos diagonalization. 
Then, the lowest eigenstates with $\langle \hat K_\lambda \rangle \approx 0$ provide the desired $| \psi_\lambda^* \rangle$. 
As noted in Ref. \cite{Long2003}, it is very difficult to find an eigenstates with $\langle \hat K_\lambda \rangle = 0$ 
for finite systems. 
Moreover, it is not guaranteed that $| \psi_\lambda^* \rangle$ are true energy eigenstates. 
Nevertheless, it will be shown that this procedure gives fairly accurate results because  
$| \psi_\lambda^* \rangle$ is dominated by the energy eigenstates whose energy eigenvalue is closest to $\lambda$. 
To obtain the target internal energy $E=\lambda$ at a given temperature $T$, 
we employ the canonical TPQ \cite{Sugiura2013} because $T$ is an input parameter. 

This paper considers the one-dimensional (1D) antiferromagnetic (AFM) spin $S=1/2$ Heisenberg model 
described by
\begin{equation}
\hat H=J \sum_{l=1}^{N} \vec S_l \cdot \vec S_{l+1}, 
\label{eq:HJ}
\end{equation}
with the periodic boundary condition, $\vec S_{N+1}=\vec S_1$. 
The nearest-neighbor coupling $J$ is taken as the unit of energy. 

Since we are interested in sizes small enough that the MCLM can be applied and numerically exact solutions are available, 
we consider relatively small size systems with a total site number $N=16$, 18, 20, and 24. 
The size of the basis, for example, for $N=20$ and 24, is $2^{20}=1,048,576$ and $2^{24}=16,777,216$, respectively. 
The full diagonalization is possible for $N=16$, 18, and 20, where we diagonalize the Hamiltonian matrix for each subspace characterized by both the total momentum $\mathbf{K}$ and the $z$ component of the total spin $S_{tot}^z$.  For $N=24$, the FTLM is used due to the exponentially increasing basis size. In the FTLM, $S(\vec q, \omega)$ is given by~\cite{Jaklic1994}
\begin{eqnarray}
S(\vec q, \omega) \!\!\!&=&\!\!\! \frac{1}{Z}\sum_s \frac{N_s}{R_s} \sum_{r_s=1}^{R_s} \sum_{i,j=0}^{M_s} e^{-\beta (\varepsilon_i^{r_s}-\varepsilon_0)} \langle r_s | \psi_i^{r_s} \rangle \nonumber \\
\!\!\!&\times&\!\!\!  
\langle \psi_i^{r_s} | \hat{X}_{\vec q} | \tilde{\psi_j^{r_s}} \rangle \langle \tilde{\psi_j^{r_s}} | \hat{X}_{-{\vec q}} | r_s \rangle \delta \left( \omega - \tilde{\varepsilon}_j^{r_s} + \varepsilon_i^{r_s} \right)
\label{eq:FTLM}
\end{eqnarray}
with the partition function
\begin{equation}
Z=\sum_s \frac{N_s}{R_s} \sum_{r_s=1}^{R_s} \sum_{i=0}^{M_s} e^{-\beta (\varepsilon_i^{r_s} -\varepsilon_0)} \left| \langle r_s | \psi_i^{r_s} \rangle \right|^2,
\label{eq:Z}
\end{equation}
where $\beta=1/T$, and
the summation $s$ runs over all possible subspaces characterized by $\mathbf{K}$ and $S_{tot}^z$; $N_s$ is the total number of basis in the subspace $s$;  
vectors $| r_s \rangle$ are initial random vectors generating approximate eigenvectors $| \psi_i^{r_s} \rangle$ with energies $\varepsilon_i^{r_s}$ 
through the Lanczos iteration with step number $M_s$ for the subspace $s$; 
approximate eigenvectors $| \tilde{\psi}_j^{r_s} \rangle$ are generated by setting $\hat{X}_{-{\vec q}} | r_s \rangle$ as initial vectors for the Lanczos iteration; 
and $\varepsilon_0$ is the ground-state energy of the whole system. We take $M_s\sim 150$ and the number of random vectors $R_s\sim 60$ for the subspaces with $S_{tot}^z=0$.

In this work, the first Lanczos iteration number $M_1$ to diagonalize $\hat K_\lambda$ and 
the second Lanczos iteration number $M_2$ to compute dynamical quantities using the CFE are taken to be $M_1 = M_2 = 300$. 
Later, we will examine the dependence on $M_1$, as 
the convergence of  $| \psi_\lambda \rangle$ by diagonalizing $\hat K_\lambda$ depends on this parameter, 
influencing excitations at small frequencies. 
Now, it is worth mentioning the computational cost of the Lanczos exact diagonalization method and the MCLM. 
Let $h$ be the number of rows of the Hamiltonian matrix $\hat H$.
Then, the computational cost of the Lanczos diagonalization is linear in $h$ and square in the number of Lanczos iterations. 
Since the Lanczos iterations is of the order of hundreds, $h$ dominates the computational cost. 
For the MCLM, the number of rows of $\hat K_\lambda$ is equivalent to $h$. 
Since $\hat K_\lambda=(\hat H - \lambda)^2$, 
the number of non-zero elements in each row of $\hat K_\lambda$ is at most two times larger than that of $\hat H$, 
i.e., $\hat K_\lambda$ is a sparse matrix as $\hat H$. 
Therefore, the computational cost of the MCLM is linear in $h$ and square in the number of Lanczos iterations $M_1$, dominated by $h$.

As a dynamical quantity, we consider the spin dynamical structure factor $S(q,\omega)$ which is defined by Eq.~(\ref{eq:sqw}) 
with $\hat X_{q} = \hat S_{q}^z = \frac{1}{\sqrt{N}}\sum_l \hat S_i^z e^{-i q r_l}$. 
Because we consider the 1D lattice, momentum is shown as a scalar. 
For the broadening parameter, we use $\eta=0.05J$. 

\begin{figure}
\begin{center}
\includegraphics[width=0.8\columnwidth, clip]{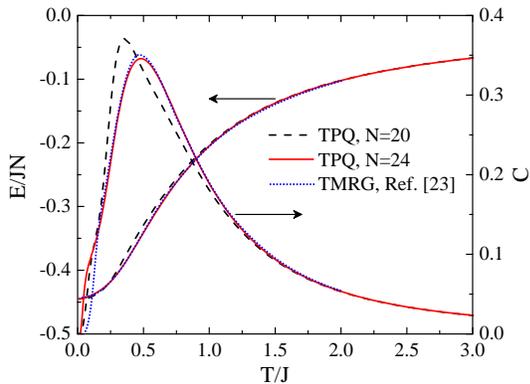}
\caption{Internal energy $E$ and specific heat $C$ of an $S=1/2$ Heisenberg chain computed by the canonical TPQ \cite{Sugiura2013}. 
Dashed (solid) curves are the results of $N= 20$ (24). 
Dotted lines are the results by a transfer-matrix renormalization-group method, corresponding to the thermodynamic limit \cite{Xiang1998}.}
\label{fig:EandC}
\end{center}
\end{figure}

\section{Results}
\label{sec:results}

Before moving into MCLM, we first check the accuracy of TPQ. 
In our implementation, 
we introduce the Taylor expansion for the canonical TPQ \cite{Sugiura2013} as
\begin{eqnarray}
|\beta \rangle = e^{-\beta \hat H/2}  |\psi_0 \rangle = \sum_{k=0} \frac{(-\beta \hat H)^k}{2^k k!} |\psi_0 \rangle, 
\end{eqnarray}
where $|\psi_0 \rangle$ is an initial vector.  
The expansion is continued until 
the $k$th-order contribution becomes negligibly small compared with the lower-order contributions. 
As mentioned in Ref.~\cite{Sugiura2013}, 
the internal energy $E = \langle \hat H \rangle$ and the specific heat $C=\{ \langle \hat H^2 \rangle - \langle \hat H \rangle^2 \}/T^2N$ as a function of $T$ 
can be computed very efficiently within a single run 
by storing only $\langle k | \hat H | k \rangle$, $\langle k | \hat H | k+1 \rangle$, $\langle k | \hat H^2 | k \rangle$, and $\langle k | \hat H^2 | k+1 \rangle$ 
with $|k\rangle = (-\hat H)^k |\psi_0\rangle$
up to the upper limit of $k=k_{max}$, which determines the lowest temperature. 
The computational cost of this canonical TPQ is, therefore, linear in $h$ and linear in $k_{max}$. 
%
For the current calculations, we used initial vectors defined by $|\psi_0 \rangle = \sum_i  c_i |i \rangle$, 
where $\{ | i \rangle \}$ is an orthonormal basis of $\hat H$ in Eq.~(\ref{eq:HJ}). 
Here, all $2^N$ states are considered, arranging Ising spins on lattice sites as $|\sigma_1^z, \sigma_2^z, \ldots , \sigma_N^z \rangle =|1,-1, \ldots, 1 \rangle $. 
We chose  for $\{ c_i \}$ a set of real numbers with uniform magnitude and random sign. 
No symmetry classifications are used to reduce the size of $\hat H$ for either the TPQ or MCLM calculations to follow. 

Numerical results for $E$ and $C$ are plotted in Fig. \ref{fig:EandC}. 
The internal energy $E$ will be used as a target energy $\lambda$ for our MCLM calculations. 
The results for $N=24$ almost recover the results by the transfer-matrix renormalization-group (TMRG) method, 
corresponding to the thermodynamic limit \cite{Xiang1998}. 
The deviation from the thermodynamic limit is enhanced at low temperatures below $T \sim J$ for $N=20$. 
For $N=24$, the deviation from the thermodynamic limit is enhanced at much lower temperatures, $T \sim 0.2 J$. 
Below these temperatures, the discreteness of the energy spectrum starts to influence thermodynamic and dynamical properties 
and, therefore, the MCLM is expected to become less accurate as TPQ. 
At much lower temperatures $T \alt 0.01J$, where the ground-state wave function is dominant, 
one anticipates that MCLM will recover the correct results, while those would be rather discrete.

\begin{figure}
\begin{center}
\includegraphics[width=1\columnwidth, clip]{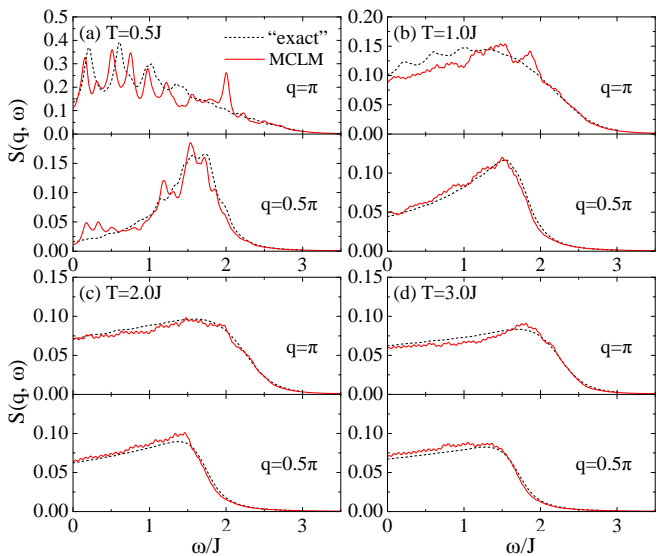}
\caption{Spin dynamical structure factor $S(q,\omega)$ of a 20-site Heisenberg chain at $q=\pi$ and $q=0.5\pi$ 
at the different temperatures indicated.
Solid lines are the results of MCLM, while dashed lines are the exact results obtained by full diagonalization of the Hamiltonian.}
\label{fig:sqw20}
\end{center}
\end{figure}

\begin{figure}
\begin{center}
\includegraphics[width=0.7\columnwidth, clip]{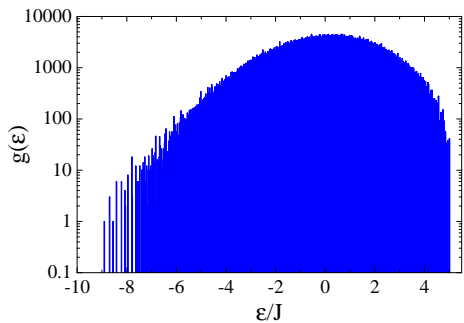}
\caption{Energy density of states $g(\varepsilon)$ of a 20-site Heisenberg chain. 
The lowest peak corresponds to the ground state with the energy $E_{GS}=-8.904 J$. 
The number of states is accumulated on an energy grid with a width $0.02J$. 
}
\label{fig:dos}
\end{center}
\end{figure}

We start from the comparison with a 20-site Heisenberg chain for which the full diagonalization is available. 
Figure \ref{fig:sqw20} shows the comparison of $S(q,\omega)$ at $q= \pi$ and $q=0.5\pi$ at the various temperatures indicated. 
It is clear to the eye that both the MCLM results and the exact results at $T=0.5J$ have a spiky structure. 
This arises from size effects. 
Peak positions are slightly different between the two calculations because excitations from different states are contributing to $S(q,\omega)$, 
namely, every state with proper Boltzmann weight in the full diagonalization results, 
while only one state with $\lambda \approx E (T=0.5J)$ in the MCLM. 
Despite this difference, the MCLM results capture the characteristic feature of the exact results quite well; 
$S(q,\omega)$ at $q=0.5 \pi$ has a peak at $\omega \sim 1.5J$--$1.7 J$, 
and such a peak is shifted to lower frequencies at around $\omega \sim 0.5J$--$J$ at $q = \pi$. 
Another point to note is that $S(q,\omega)$ by the MCLM has more peaks than those obtained by the full diagonalization, i.e., the exact result. 
This is related to the energy histogram as discussed later. 

With increasing temperature, the spiky structure diminishes rapidly and spectral shapes become a smooth continua extended down to zero frequency 
with a high-frequency shoulder at $\omega \approx 1.5 J$ at $q = 0.5 \pi$ and $\omega \approx 2 J$ at $q = \pi$. 
The spectra consist of small oscillations because of relatively small $M_2$ 
(Ref.~\cite{Prelovsek2013} argued  that $M_2$ should be $\sim 1000$). 
The exact results are reproduced by the MCLM extremely well, even though only one energy eigenstate is used. 
This comes from the fact that the energy eigenstate distribution is very dense around $\lambda$ at high temperatures, 
and therefore $S(q,\omega)$ consists of dense poles. 

It is instructive to check the energy density of states (DOS) $g(\varepsilon)$ for the $N=20$ case because all the energy eigenvalues are available. 
For larger systems, the recursion method could be used to compute the energy DOS and thermodynamic properties directly from it \cite{Otsuka1995}.
As shown in Fig.~\ref{fig:dos}, $g(\varepsilon)$ is continuous in energy $\varepsilon$ in the high-energy regime $\varepsilon-E_{GS} \agt J$, 
where $E_{GS}=-8.904 J$ is the ground-state energy. 
At lower energy $\varepsilon-E_{GS} \alt J$, the energy DOS is discrete and the peaks in $g(\varepsilon)$ are separated. 
Therefore, one expects that $S(q,\omega)$ starts to show a discrete structure at $T \alt J$, 
which explains the spiky feature in Fig.~\ref{fig:sqw20} (a). 
Note that in addition to the discreteness or continuity of the energy DOS, 
the details of operators $\hat X_q$ could induce frequency and momentum-dependent features in $S(q,\omega)$. 
As noted earlier, $S(q,\omega)$ by the MCLM has more peaks than that by the full diagonalization. 
This could be understood based on $g(\varepsilon)$. 
Our TPQ calculation for $N=20$ gave the internal energy $E= -6.582 J$ at $T=0.5J$. 
Thus, the $| \psi_\lambda \rangle$ picked up by the MCLM is in the continuum of $g(\varepsilon)$. 
This is because the internal energy is given by 
$E=\int d\varepsilon \, \varepsilon \exp(-\varepsilon/T) g(\varepsilon)/\int d\varepsilon \exp(-\varepsilon/T) g(\varepsilon)$, 
and $g(\varepsilon)$ is an exponentially increasing function of $\varepsilon$ at $\varepsilon-E_{GS} \alt 8J$. 
Note that the energy distribution function $\rho(\varepsilon)$ and $g(\varepsilon)$ are related by $\rho(\varepsilon)=\exp(-\varepsilon/T) g(\varepsilon)$. 
As a result, $S(q,\omega)$ obtained by the MCLM suffers from the discreteness of $g(\varepsilon)$ at $\varepsilon-E_{GS} \alt J$ less than that by the full diagonalization, although the MCLM is not exact in small systems. 
While it is difficult to calculate the full $g(\varepsilon)$ for large systems, 
from the $N=20$ example 
it is expected that the MCLM could provide continuous dynamical response functions in wider temperature regimes or lower temperatures than the FTLM 
because the lowest energy, above which $g(\varepsilon)$ is continuous, is expected to lower with increasing $N$. 

\begin{figure}
\begin{center}
\includegraphics[width=1\columnwidth, clip]{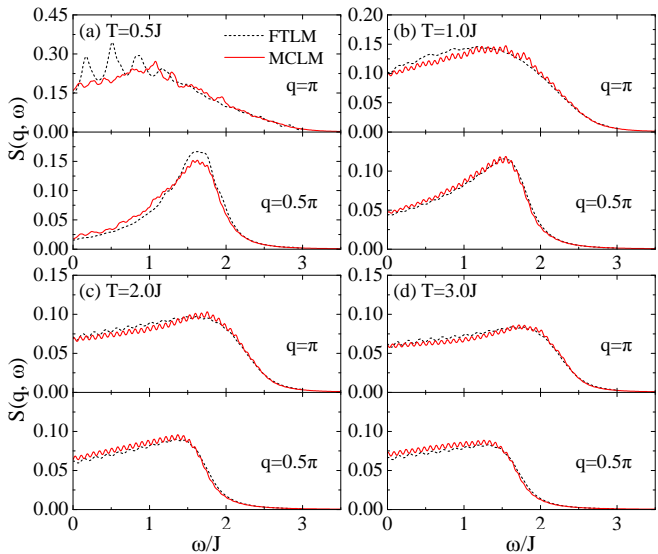}
\caption{Spin dynamical structure factor $S(q,\omega)$ of a 24-site Heisenberg chain at $q=\pi$ and $q=0.5\pi$ 
at the different temperatures indicated.
Solid lines are the results of MCLM, while dashed lines are the results of FTLM.}
\label{fig:sqw24}
\end{center}
\end{figure}

We now turn to the larger system size $N=24$. 
Figure \ref{fig:sqw24} shows the comparison of $S(q,\omega)$ at $q= \pi$ and $q=0.5\pi$ at the various temperatures indicated. 
At $T=0.5J$, the FTLM result displays a spiky structure, similar to the full diagonalization result for $N=20$. 
While the MCLM result at the same temperature showed a similar spiky structure for $N=20$, 
such a structure is greatly suppressed for $N=24$ because of the dense distribution of energy eigenstates around $\lambda$. 
This indicates that size effects at finite temperatures are smaller for the MCLM than for the full diagonalization or for the FTLM.

\begin{figure}
\begin{center}
\includegraphics[width=0.8\columnwidth, clip]{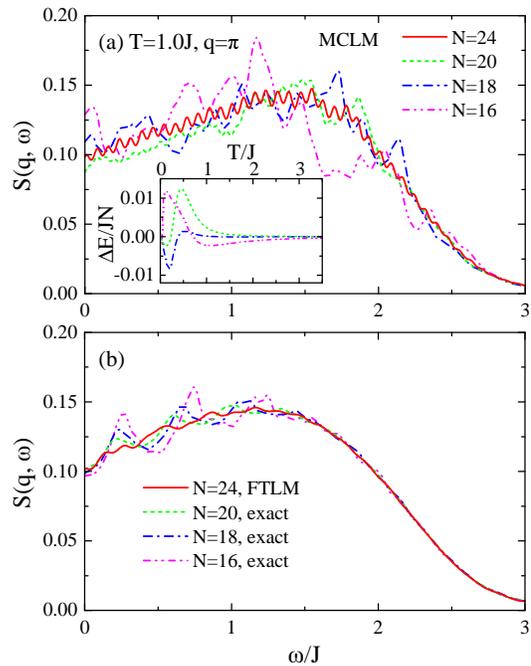}
\caption{Comparison of spin dynamical structure factor $S(q,\omega)$ of 16-, 18-, 20-, and 24-site Heisenberg chains at $T=J$ and $q=\pi$. 
The results of MCLM are shown in (a), and 
the result of FTLM ($N=24$) and the numerically exact results of $N=16$, 18, and 20 are shown in (b). 
The inset shows the TPQ results of the internal energies of $N=16$, 18, and 20 measured from that of $N=24$. }
\label{fig:smallsizes}
\end{center}
\end{figure}

While MCLM could become accurate for large systems, its accuracy could be lost rapidly for small systems 
because the inequivalence between the microcanonical ensemble and the canonical ensemble grows as the system size is decreased. 
To see how the MCLM loses its accuracy, we consider smaller systems, $N=16$ and 18. 
The internal energy $E=\lambda$ is computed using the canonical TPQ, as mentioned earlier. 
The inset of Fig.~\ref{fig:smallsizes} (a) shows the internal energy of $N=20$, 18, and 16, measured from that of $N=24$, 
$\Delta E/N \equiv E/N - E(N=24)/24$ . 
Even for the smallest system considered here, the deviation of $E/N$ is at the largest $\sim 0.01J$, which is roughly $2\%$ of the internal energy. 
Thus, for this calculation, the size effect in the canonical TPQ is rather small. 
However, the size effect is amplified in the specific heat and entropy because of the temperature derivative. 

The main panels of Fig.~\ref{fig:smallsizes} compare $S(q=\pi,\omega)$ of 16-, 18-, 20-, and 24-site Heisenberg chains at $T=J$; 
the MCLM results in (a), 
and the FTLM result ($N=24)$ and the numerically exact results of $N=16$, 18, and 20 in (b). 
With decreasing $N$, $S(q=\pi,\omega)$ by MCLM develops a spiky structure rapidly. 
A similar behavior is also seen in (b), 
but the difference between different $N$'s is rather small because of the thermal averaging over different eigenstates $| \psi_i \rangle$ with proper weights $e^{-\varepsilon_i/T}$. 
On the other hand, the MCLM does not involve such thermal averaging, and, therefore,  
the discreteness in the energy spectra largely influences dynamical quantities in small systems as one could anticipate.

\begin{figure}
\begin{center}
\includegraphics[width=0.8\columnwidth, clip]{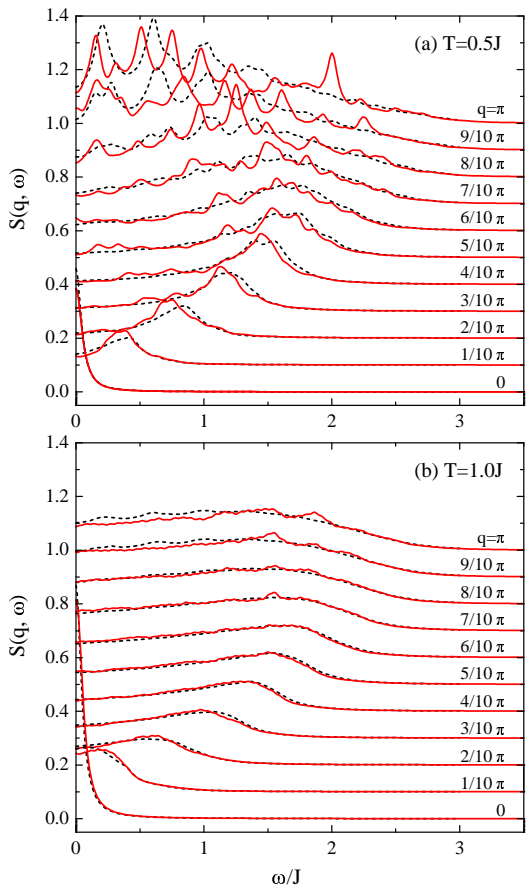}
\caption{Spin dynamical structure factor $S(q,\omega)$ of a 20-site Heisenberg chain at $q=n \pi /10$ with $n=0$--10 
at the temperatures indicated.
Solid lines are the results of MCLM, and dash lines are the exact results. 
Data with different values of $q$ are shifted vertically by 0.1 for clarity. 
}
\label{fig:sqw20Q}
\end{center}
\end{figure}

\begin{figure}
\begin{center}
\includegraphics[width=0.8\columnwidth, clip]{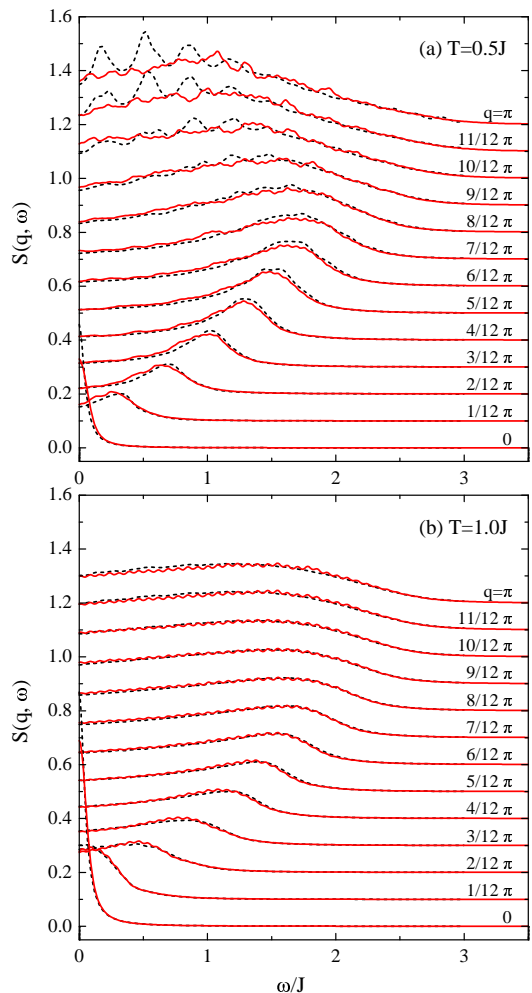}
\caption{Spin dynamical structure factor $S(q,\omega)$ of a 24-site Heisenberg chain at $q=n \pi /12$ with $n=0$--12 
at the temperatures indicated.
Solid lines are the results of MCLM, while dashed lines are the results of FTLM. 
Data with different values of $q$ are shifted vertically by 0.1 for clarity.
}
\label{fig:sqw24Q}
\end{center}
\end{figure}

Figures \ref{fig:sqw20Q} and \ref{fig:sqw24Q} summarize $S(q,\omega)$ at all momenta with different temperatures for $N=20$ and for $N=24$, 
respectively. 
In spite of the spiky structure, the MCLM results at $T=0.5J$ reproduce the exact results fairly well, 
especially the broad peak position as a function of $q$.  
The overall agreement between the MCLM and the exact results or the FTLM results is excellent at $T=J$. 

\begin{figure}
\begin{center}
\includegraphics[width=0.8\columnwidth, clip]{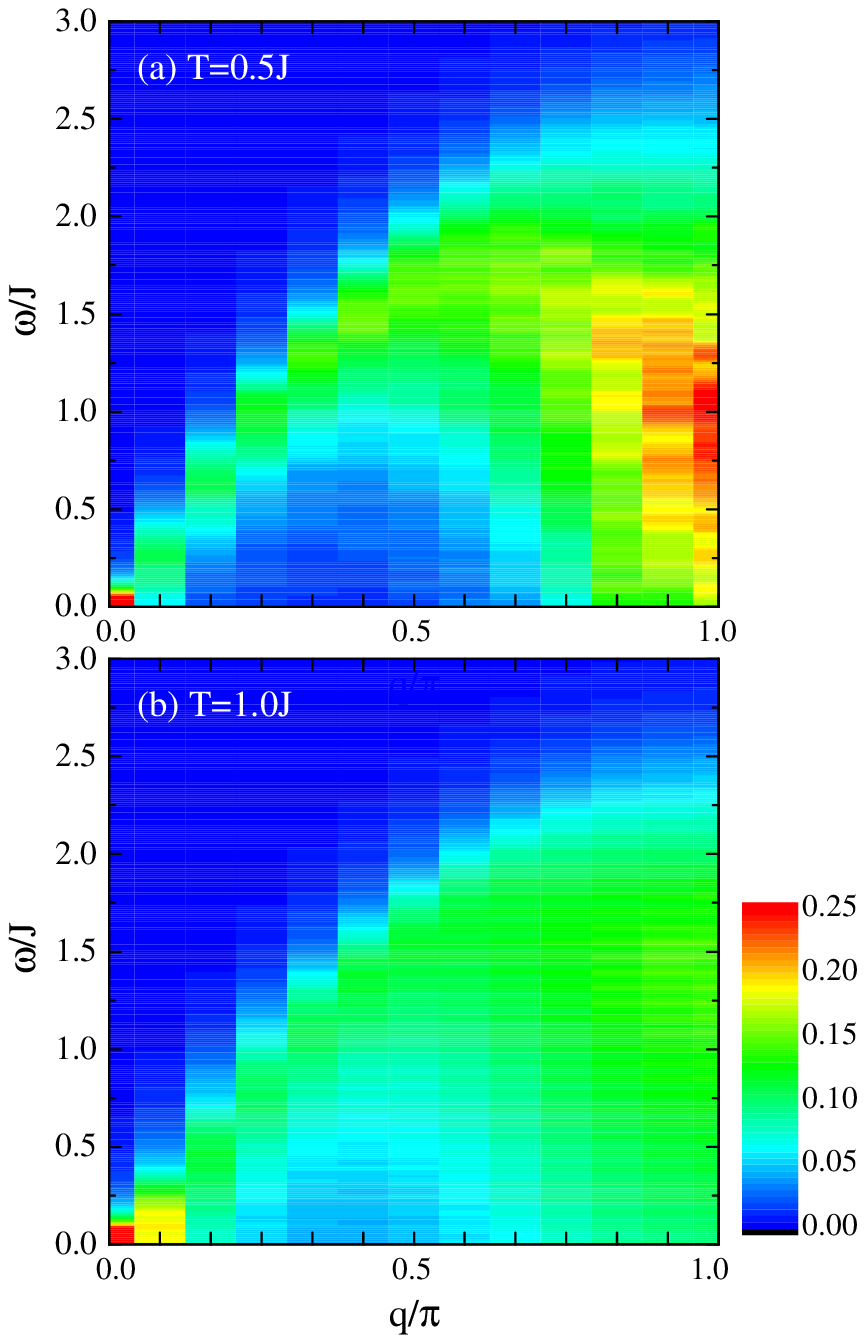}
\caption{Contour plots of the spin dynamical structure factor $S(q,\omega)$ of a 24-site Heisenberg chain computed by the MCLM 
at the temperatures indicated.
}
\label{fig:sqw24Qcont}
\end{center}
\end{figure}

While already observed for $N=20$, it is clearer for $N=24$ that 
the position of the spectral intensity changes rather drastically between $T=0.5J$ and $T=J$; 
$S(q,\omega)$ has a broad maximum at $q \sim \pi$ and $\omega \sim J$ at $T=0.5J$, 
while it has a sharp maximum at $q \sim 0$ and $\omega \sim 0$ at $T=J$. 
Note that the highest peak is always located at $q = 0$ and $\omega =0$ at all temperatures investigated in this work, 
and the peak intensity increases with increasing temperature. 
This behavior is seen in contour plots in Fig. \ref{fig:sqw24Qcont} more clearly. 
At $T=0.5J$, one can see a remnant of a two-spinon continuum, bounded by 
$\varepsilon_U = \pi J |\sin \frac{q}{2}|$ and $\varepsilon_L = \frac{\pi J}{2} |\sin q|$ at $T=0$ 
but shifted toward lower frequencies at finite temperatures. 
At $T=J$, the lower bound for the continuum is only visible at $q \sim \pi/2$. 
These results well reproduce previous reports using the QMC methods \cite{Staykh1997,Grossjohann2009} 
and the finite temperature DMRG method \cite{Barthel2009}, 
despite the smaller size and computational simplicity in the current study. 

\begin{figure}
\begin{center}
\includegraphics[width=0.8\columnwidth, clip]{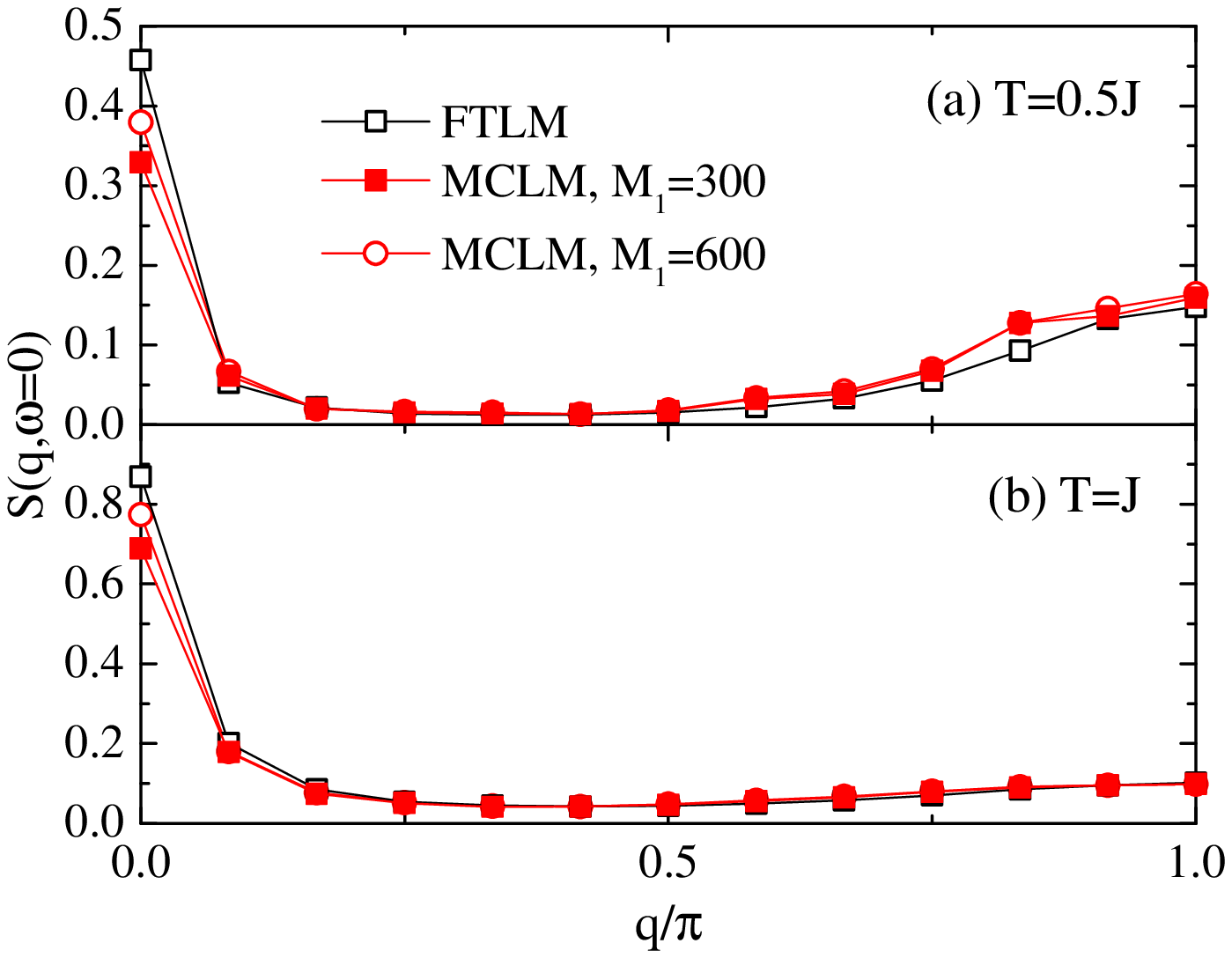}
\caption{Spin dynamical structure factor $S(q,\omega=0)$ of a 24-site Heisenberg chain as a function of $q$ at 
$T=0.5J$ (a) and $T=J$ (b). 
Filled squares are the results of the MCLM with $M_1=300$, while open squares are the results of the FTLM. 
Open circles are the results of the MCLM with $M_1=600$.
}
\label{fig:sq0}
\end{center}
\end{figure}

As seen in Figs. \ref{fig:sqw20Q} and \ref{fig:sqw24Q}, $S(q,\omega)$ at $q=0$ has a single peak at $\omega=0$ 
corresponding to the short-range diffusive ferromagnetic correlation \cite{Staykh1997}. 
The peak height obtained by the MCLM is always smaller than that by the FTLM. 
Here, we check the results at the $\omega=0$ limit in more detail. 
Figure \ref{fig:sq0} compares the $S(q,\omega=0)$ of a 24-site Heisenberg chain at $T=0.5J$ (a) and $T=J$ (b). 
The overall $q$ dependence by the MCLM or by the FTLM is consistent with that by the QMC results obtained for larger systems reported in Ref.~\cite{Staykh1997}. 
However, the peak height at $q=0$ and $q \sim \pi$ at $T=0.5J$ and $q=0$ at $T=J$ by the MCLM or by the FTLM is 
about 10--20~\% smaller than the QMC results. 
In addition to the size effects and the finite broadening $\eta$, 
the MCLM results in the $\omega=0$ limit could have been underestimated by the relatively small $M_1$. 
As discussed in Ref.~\cite{Long2003}, 
the $M_1=300$ used in the current work may not be large enough to resolve the small energy resolution near the target internal energy $E=\lambda$, 
but this could be systematically improved by increasing $M_1$. 
As shown by the open circles in Fig.~\ref{fig:sq0}, results with $M_1=600$ improve the low-energy behavior peaked at $q=0$.

\begin{figure}
\begin{center}
\includegraphics[width=0.8\columnwidth, clip]{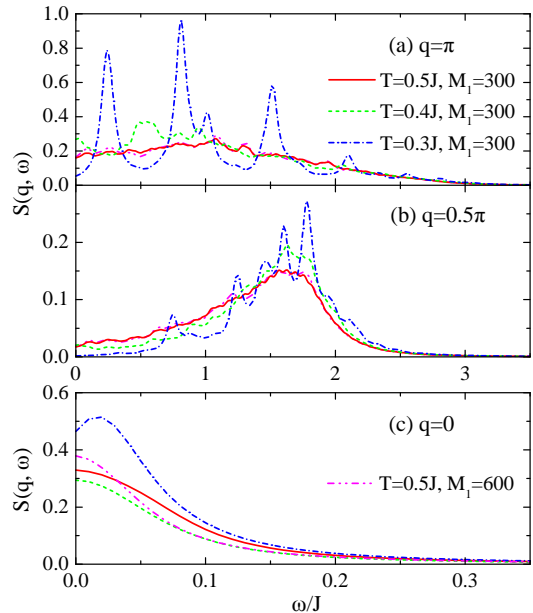}
\caption{Spin dynamical structure factor $S(q,\omega)$ of a 24-site Heisenberg chain at $q=\pi$ (a), $q=0.5\pi$ (b), and $q=0$ (c) 
at low temperatures computed by the MCLM with $M_1=300$. 
Solid, dashed, and dash-dot lines are the results at $T=0.5J$, $0.4J$, and $0.3J$, respectively. 
Dash-dot-dot lines are the results at $T=0.5J$ with $M_1=600$.
}
\label{fig:sqwT}
\end{center}
\end{figure}

Finally, we examine the low-temperature behavior of $S(q,\omega)$ to analyze the low-temperature limit where the MCLM could provide 
a continuous spectra as a function of frequency, which is closely related to the separation between the energy eigenstates near the target internal energy. 
Figure \ref{fig:sqwT} shows $S(q,\omega)$ at $q=\pi$ and $0.5\pi$ computed by the MCLM with $M_1=300$ in the low-temperature regime. 
The spectrum at $q=\pi$ starts to develop a peak structure at $T=0.4J$, 
but it is less pronounced than the similar structure produced by the FTLM at $T=0.5J$ 
[see Fig. \ref{fig:sqw24Q} (a)]. 
At $T=0.3J$, the spectra at $q=\pi$ and $q=0.5\pi$ are clearly dominated by several peaks, which resemble those in the FTLM at $T=0.5J$. 
Furthermore, the spectrum at $q=0$ starts to increase and the peak position shifts from $\omega=0$ at $T=0.3J$. 
These indicate that the energy spectrum is discrete around the target internal energy $E(T=0.3J)$. 
Since the energy spectrum depends on the model under consideration, 
a general statement cannot be made. 
But at least for the 1D Heisenberg model and considering the lowest temperature at which continuous spectral functions appear, 
$T \agt 0.4J$ for the MSLM vs $T \agt 0.5J$ for the FTLM, 
the MCLM could provide continuous spectral functions at $\approx 20 \%$ lower temperatures than the FTLM. 
This might indicate that the MCLM is more suitable than the FTLM to {\it visualize} the dynamical properties of an interacting quantum model 
in the thermodynamic limit in a wider range of temperature. 
Figure \ref{fig:sqwT} also shows the results with $M_1=600$ at $T=0.5J$ (dash-dot-dot lines). 
While the results with $M_1=300$ and $M_1=600$ are almost indistinguishable at $q=\pi$ and $q=0.5 \pi$, 
increasing $M_1$ does change the results, especially at the low-energy peak at $q=0$. 
This is because $M_1$ controls the resolution of energy eigenstates near $\lambda$. 

For electronic models, such as Hubbard, $tJ$, and Anderson models, 
smooth dynamical spectral functions could be achieved by introducing twisted boundary conditions (BCs) \cite{Tsutsui1996,Tohyama2004} 
and averaging over different BCs. 
Such boundary averaging could widen the temperature range for which both MCLM and FTLM could be utilized. 

\begin{figure}
\begin{center}
\includegraphics[width=0.8\columnwidth, clip]{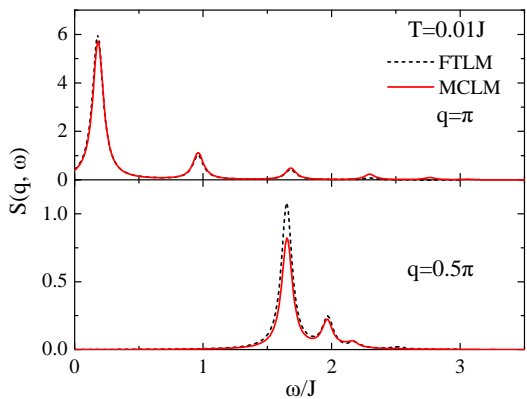}
\caption{Spin dynamical structure factor $S(q,\omega)$ of a 24-site Heisenberg chain at $q=\pi$ and $q=0.5\pi$ at $T=0.01J$. 
Solid lines are the results of MCLM with $M_1=300$, while dashed lines are the results of FTLM. 
}
\label{fig:sqw001}
\end{center}
\end{figure}

Far below $T=0.3J$, the MCLM and the FTLM provide consistent results in a regime where $S(q,\omega)$ is dominated by discrete levels. 
As shown in Fig. \ref{fig:sqw001}, the peak position of $S(q,\omega)$ is indistinguishable between the two methods, 
although the peak height is slightly different because different energy eigenstates contribute differently. 
At much lower temperatures, the results by the two methods completely agree. 

\section{Summary and discussion}
\label{sec:summary}

To summarize, we have examined the applicability of the MCLM developed in Ref.~\cite{Long2003} 
to finite-size systems. 
While the microcanonical ensemble is supposed to provide the same results as the canonical ensemble, becoming exact in the thermodynamic limit, 
we have found that the MCLM provides reasonably accurate results even for rather small systems. 
The computational cost for the whole process described in this paper is similar to that of the zero-temperature Lanczos calculations. 
The target internal energy $E = \lambda$ at a given temperature is computed using a TPQ method developed in Ref.~\cite{Sugiura2013}. 
This does not require diagonalizing a Hamiltonian matrix $\hat H$ at all. 
For a given temperature, the diagonalization is done only once or a finite number of times using a new operator $\hat K_\lambda= (\hat H -\lambda)^2$ 
to obtain (approximate) energy eigenstates $|\psi_\lambda \rangle$ with the corresponding energy $\lambda$. 
This contrasts with the FTLM, which requires the diagonalization process multiple times to carry out the statistical average. 
Such a numerical inexpensiveness suggests that the MCLM could serve as a useful tool to analyze the dynamical properties of 
interacting quantum models at finite temperatures. 
The MCLM should outperform the FTLM if not too many temperature points are needed, 
because the FTLM could obtain all temperature results by a single calculation. 
Furthermore, the MCLM provides continuous spectral functions, representing the properties in the thermodynamic limit, 
in a wider temperature range than the FTLM. 

Since the microcanonical ensemble is used in the MCLM, 
one might ask if the microcanonical TPQ $| \psi_k \rangle \propto (E_{max} - \hat H)^k | 0 \rangle$ 
could be used as the energy eigenstate $| \psi_\lambda \rangle$. 
Here, $E_{max}$ is the largest eigenenergy of $\hat H$ and $| 0 \rangle$ is a random initial vector, 
and the internal energy $E$ and temperature $T$ are determined by the prescription described in Ref. \cite{Sugiura2012}. 
For very large systems, this might be a good strategy. 
However, for small systems $| \psi_k \rangle$ has nonzero contributions from other energy eigenstates than $| \psi_\lambda \rangle$. 
Therefore, it is not clear how the off-diagonal elements of $\hat H$ influence the physical quantities. 

The canonical TPQ $| \beta \rangle$ \cite{Sugiura2013} 
is not an energy eigenstate, 
and therefore it cannot directly replace $|\psi_\lambda \rangle$ in Eq. (\ref{eq:sqw}), either. 
Nevertheless, we found it formally possible to express a dynamical quantity by extending the canonical TPQ 
by doubling the basis set as $| \beta \rangle = \sum_i a_i | i \rangle \Rightarrow |\tilde \beta \rangle = \sum_{i,j} a_i \delta_{ij} | i \rangle \otimes |j\rangle$ 
and introducing the Liouville operator defined by $\hat L \equiv \hat H \otimes \hat I - \hat I \otimes \hat H$, 
where operators on the left (right) side of $\otimes$ act only in the physical (auxiliary) space described by the first (second) index 
$i$ ($j$) in $ |\tilde \beta \rangle$. 
Such basis (Hilbert space) doubling is also used for finite-temperature DMRG methods \cite{Feiguin2005,Tiegel2014}, 
and for solving a quantum master equation for the nonequilibrium dynamical mean field theory \cite{Arrigoni2013}. 
With the extended canonical TPQ and the Liouville operator, 
a dynamic quantity is expressed as 
\begin{equation}
S(\vec q, \omega)
= - 
\frac{1}{\pi Z} \, {\rm Im}
\langle \tilde \beta | \hat X_{\vec q} \otimes \hat I  \Bigl[\omega+i \eta  - \hat L \Bigr]^{-1} \! \hat X_{-\vec q} \otimes \hat I | \tilde \beta \rangle ,
\label{eq:Liouville} 
\end{equation}
with $Z = \langle \beta | \beta \rangle = \langle \tilde \beta | \tilde \beta \rangle$. 
Starting from $\hat X_{-\vec q} \otimes \hat I | \tilde \beta \rangle$, 
one could formally use the CFE with $\hat L$ instead of $\hat H$. 
It is straightforward to see that Eq. (\ref{eq:Liouville}) recovers the correct expression by realizing 
$| \tilde \beta \rangle \propto \sum_{m,n} e^{-\varepsilon_m/2 T} \delta_{mn}| \psi_m \rangle \otimes | \psi_n \rangle$, 
with $|\psi_m \rangle$ being an energy eigenstate with the eigenvalue $\varepsilon_m$. 
However, because of the basis doubling, the required computer memory size grows faster than the TPQ. 
Therefore, the use of this method is limited to very small systems for which the full diagonalization would be possible, 
unless one finds a clever way to truncate the basis states describing $\hat L^k \hat X_{-\vec q} \otimes \hat I  | \tilde \beta \rangle$ with $k=0,1,2, \ldots$

Recently, we became aware of a similar but distinct approach to compute finite-temperature dynamical quantities of interacting quantum models based on  a shifted Krylov subspace method \cite{Yamaji2018}. 
Similar to FTLM, this method is based on the canonical ensemble and samples excited states with proper Boltzmann weights. 
The excited eigenstates are filtered from TPQ by the filter operator. 
While the computational algorithm is different and the cost could be more expensive than the MCLM in general, 
this method is supposed to become equivalent to the current MCLM approach when the single excited eigenstate $| \psi_\lambda \rangle$ is filtered.

Having examined the applicability of the MCML, the system size is still limited by the exponentially growing basis states. 
In order to apply a numerical diagonalization method based on the microcanonical ensemble for larger systems, 
it would be necessary to truncate the basis states systematically. 
One possible direction would be applying DMRG methods as follows: 
1. Perform the static calculation at temperature $T$ to compute the target internal energy $E=\lambda$ of the system under investigation. 
2. Optimize the many-body wave function $| \psi_\lambda \rangle$ targeting the energy $\lambda$ using $\hat K_\lambda = (\hat H - \lambda)^2$ 
or using the DMRG-X algorithm \cite{Khemani2016,Devakul2017}. 
and 3. Compute dynamical quantities using $| \psi_\lambda \rangle$. 
By this procedure, one might be able to increase the system size within the limit of computer resources. 
This work will be pursued in the near future. 

\acknowledgements
The research by S.O. and G.A. was supported by the Scientific Discovery through Advanced Computing (SciDAC) program 
funded by the U.S. Department of Energy, Office of Science, Advanced Scientific Computing Research and Basic Energy Sciences, 
Division of Materials Sciences and Engineering.
The research by E.D. is supported by the  U.S. Department of Energy, Office of Science, Basic Energy Sciences, Materials Sciences and Engineering Division. 
T.T. is supported by MEXT, Japan, as a social and scientific priority issue (creation of new functional
devices and high-performance materials to support next-generation industries) to be tackled by using a
post-K computer. 
%
We thank P. Prelov{\v s}ek, X. Zotos, and Y. Yamaji for their helpful discussions.

\end{document}